\documentclass[sigconf, 10pt, nonacm=true]{acmart}
\usepackage{xspace}
\usepackage{booktabs}
\usepackage{multirow}

\graphicspath{{./figs/}}

\providecommand{\tun}{{\tt tun}\xspace}
\providecommand{\m}{{\tt MBZ}\xspace}

\newcommand{\ednote}[2]{}

\newenvironment{itemize-s}%
  {\begin{itemize}%
    \setlength{\itemsep}{0pt}%
    \setlength{\parskip}{0pt}}%
  {\end{itemize}}
  
\newcommand{\ie}{{\it i.e.,}\xspace}
\newcommand{\eg}{{\it e.g.,}\xspace}
\newcommand{\Eg}{{\it E.g.,}\xspace}

\DeclareGraphicsExtensions{.pdf} 

\newcommand{\xref}[1]{\S~\ref{#1}}
\newcommand{\parax}[1]{\noindent \textbf{#1:}}

\setcopyright{none}
\acmConference[arXiv]{}
\acmYear{}
\copyrightyear{}
\acmISBN{}
\acmDOI{}

\begin{document}

\title[Back in Control]{Back in control --- An extensible middle-box on your phone}
\author{James Newman$^*$\hspace{10pt}Abbas Razaghpanah$^+$\hspace{10pt}Narseo Vallina-Rodriguez$^\dag\ddag$\\ Fabi\'an E. Bustamante$^*$\hspace{10pt}Mark Allman$^\ddag$\hspace{10pt}Diego Perino$^\diamond$\hspace{10pt}Alessandro Finamore$^\diamond$}
\affiliation{Northwestern University$^*$ Stony Brook University$^+$ IMDEA Networks Institute$^\dag$
ICSI$^\ddag$ Telefonica Research$^\diamond$}

\renewcommand{\shortauthors}{J. Newman et al.}

\begin{abstract}


The closed design of mobile devices --- with the increased security and consistent 
user interfaces--- is in large part responsible for their becoming the 
dominant platform for accessing the Internet.  These benefits, however, are
not without a cost. Their operation of mobile devices and their apps is not
easy to understand by either users or operators.

We argue for recovering transparency and control on mobile devices
through an extensible platform that can intercept and modify traffic before leaving the 
device or, on arrival, before it reaches the operating system. Conceptually, this is the 
same view of the traffic that a traditional middlebox would have at the far end of the 
first link in the network path. We call this platform ``middlebox zero'' or \m. By being 
on-board, \m also leverages local context as it processes the traffic and complement 
the network wide view of standard middleboxes.  We discuss the challenges of the \m 
approach, sketch a working design, and illustrate its potential with some concrete 
examples.


\end{abstract}

\maketitle
\thispagestyle{empty}

\section{Introduction}
\label{sec:intro}

Mobile devices are indispensably handy and fundamentally enigmatic.
In just over ten years, they have become the dominant platform
for accessing the Internet. This is at least partially due to their closed
nature, a design decision clearly stated by one of its main designers --- 
Steve Jobs ---``We define everything that is on the phone''~\cite{jobs:nytimes07}.

This closed design, and a policed environment where the manufacturer has
veto power over every application, is clearly beneficial for security and
ensures user-friendly, consistent interfaces. These benefits are
not without costs. 

The operation of mobile devices and apps is hard to 
understand. Noticeable changes in performance are as commonplace as
difficult to diagnose and even basic questions, such as \emph{why is this
website not loading?} or \emph{with what other sites is this website communicating?} 
do not have ready answers. 

This challenge has inspired different approaches to return visibility and
control to users and developers. Some tools will let users change the 
behavior of applications without accessing the APK~\cite{www:xposed} or 
installing a BusyBox-like toolset~\cite{www:busybox}. Most of them, however, 
require users to ``root'' their phones. A 2014 survey with over
14k users found that 63\% have rooted their primary Android
device~\cite{www:rootpoll}. Unfortunately, rooting a phone comes with its
own risks. Beyond the fact that this may void a device's warranty or
result on the device being ``bricked'', rooting has been linked to mobile 
malware attacks and adware~\cite{www:adware}. In a 2014 
announcement,\footnote{https://www.gartner.com/newsroom/id/2753017}
Gartner predicted that 75\% of mobile security incidents will be due to
mobile application misconfiguration with the biggest threat being devices
altered at an administration level. 

VPN APIs on mobile devices offer another option to regain control
over application's traffic. VPN APIs enable the development of mobile 
apps, mostly for improved security and privacy, without requiring root
access thus being easy for users to adopt while eliminating/reducing the 
risk of attacks. Today, however, all apps relying on it are custom, mutually
exclusive solutions to specific problems. 

Choffnes~\cite{choffnes2016case} recently proposed creating Personal Virtual Networks to
give users the illusion of a home network no matter what the network they are actually 
connected to. This paradigm, similar to typical VPN use-cases, addresses the need from 
users to gain back control over their own traffic when on untrusted networks.
The VPN server, fully owned and controlled by the end-user, can implement different 
policies for network traffic. However, user traffic is still transported to a remote 
server which lacks access to potentially valuable end-user and device context.

On the network operator side the approach to control and optimize mobile
traffic involves placing middleboxes in the network path. Middleboxes are
so popular that many enterprise networks now have as many middleboxes as
routers~\cite{sherry2012making} serving as firewalls, performance
enhancing proxies, NATs, or deep packet inspectors. The advent of Network
Function Virtualization (NFV) significantly reduced middlebox deployment
cost and time, while current deployment of edge computing infrastructures
enables the placement of middlebox-like functions very close to the 
end-user device. While clearly powerful and with an unparalleled network-wide
view, middleboxes still show limitations in terms of scalability,
especially in presence of mobile users. In addition, as in the case of Personal 
Virtual Networks, these middleboxes also lack access to potentially relevant context.

{\em We argue for recovering transparency and control on mobile devices through an 
extensible platform that can intercept and modify traffic before leaving the 
device or, on arrival, before the traffic is pass on to the operating system.}
 
Conceptually, this is the same view of the traffic that a traditional middlebox would 
have at the far end of the first link in the network path and so we call this
approach ``middlebox zero'' or \m. By existing on-board, \m can leverage 
local context to process the traffic and complement the network-wide view and optimization 
capabilities of traditional middleboxes. Rather than a dedicated middlebox, \m can 
be extended with user-specified extensions that manipulate traffic for
transparency or control. This approach has been used before to monitor
network performance~\cite{wu2016mopeye,shuba2015antmonitor} and privacy leaks~\cite{razaghpanah2015haystack}, however,
\m takes the idea a step further. Instead of having standalone,
custom-built solutions for each problem, the extensible nature of \m
allows the user to install multiple, different plugins side-by-side. 

\m opens up a wide range of opportunities and interesting challenges: from the 
architecture of an \m that could support dynamic extensibility, with minimum 
overhead and without compromising security, to new forms of interactions
between \m and traditional middleboxes an \m'ed device can run into as it moves 
between networks. The \m extensibility could also enable a service marketplace, 
wherein third parties can advertise new extensions since no single provider could 
expect to offer all imaginable needs. This alone brings a plethora of research 
questions, from how to declare extensions to the most appropriate model for their 
management and control.



\section{Motivation and background}
\label{sec:motivation}

We now motivate the need for \m by first examining the benefits 
and limitations of existing solutions and related work to control and optimize 
mobile traffic. We argue that full control and optimization can best be achieved 
via an extensible middlebox-like platform on the end-user device, capable of 
collaborating with in-network middleboxes.

\subsection{User Device Solutions}
\label{sec:rooting}

When faced with unanswerable questions regarding device behavior or
network activity, a popular approach is rooting the device.
This returns control to the device's owner and allows them to install 
unlicensed applications or gain access to the full Linux
kernel with command such as \emph{tcpdump} for network debugging. 
To root their device, most users turn to rooting tools such as
KingoRoot~\cite{www:kingo} or SuperSu\cite{www:supersu}. While these tools
streamline the rooting process, they require technically skilled users, 
and often the development of custom commands and applications to control 
and optimize traffic. More problematically, these tools have been linked
to malware \cite{www:malware} and adware~\cite{www:adware} attacks.

Recently, native support of VPN APIs by the main mobile platforms
allowed developers to capture and manipulate traffic generated by
applications without root access. This has led to the development of
applications for privacy control, measurements, and firewalls such as
Lumen~\cite{razaghpanah2015haystack},
AntMonitor~\cite{shuba2015antmonitor}, Mopeye~\cite{wu2016mopeye} or
NetGuard~\cite{www:netguard}, that can be easily installed and effectively
achieve their goals without exposing users to potential attacks. 

We see these apps as first steps towards on-board functions for traffic 
control and optimization, showing growing users' interest. Each of them, 
however, have been developed and optimized --- from scratch --- 
for a specific function and cannot run concurrently with other similar 
apps.~\footnote{Android only allows one VPN application at the time
for security reasons. The right is revoked when another application is
granted the VPN permission~\cite{www:vpnservice}.}
We argue for an extensible platform supporting multiple functions in parallel,
each of which can be more easily developed by reusing core traffic
manipulation code (i.e., an on-board middlebox), and instantiated on demand.


\subsection{Network Middleboxes}
\label{sec:middlebox}


For companies and ISPs that do not have access to the end device, the
most suitable option to control and optimize mobile traffic is to deploy 
a middlebox in the network. This allows them to gain access to all the 
traffic generated from any device connected through the middlebox. 
Middleboxes provide several services, from firewalls to protocol accelerators. 
For instance, proxies have been used to enable 
mobile data compression on the fly~\cite{agababov2015flywheel}, and previous 
work focused on how protocols and middleboxes interact~\cite{medina2004measuring,craven2014middlebox}, 
how middleboxes can be designed to enable innovation~\cite{sekar2011middlebox}, 
and how devices can establish their own private network~\cite{choffnes2016case}. 


With the advent of NFV and growing popularity of cloud platforms, 
middleboxes can be developed in software and instantiated on-demand using 
commodity hardware already installed in the
network~\cite{chen2015virtual,sherry2012making,lan2016embark,lu2016storm,qazi2013simple,ledjiar2017network}, drastically reducing deployment cost and time. To improve cloud middlebox deployment, 
several projects studied how middleboxes can be consolidated~\cite{sekar2012design}, 
or made extensible~\cite{anderson2012xomb}. The current deployment of edge computing 
infrastructures enables the placement of middlebox-like functions very close to users' devices, 
in Central Offices or even in eNodeBs. This provides middleboxes a network-wide view and the 
opportunity to jointly optimize aggregate traffic of multiple users, e.g., content placement, 
caching, network routing. However, they still have limitations that can only be solved by 
pushing some of middlebox services and features to the end device.

First, despite highly distributed and optimized NFV systems, scalability issues may arise if 
all users offload all their desired functions to the network.  
Second, middleboxes lack user context that exists on each end device.
Because the middlebox sits within the network, it is for instance difficult to
understand if traffic generated by a device was initiated by 
a given application on that device due to a user action or if it
was generated in the background. Furthermore, private
information still leaves the user device and is transferred and processed in
the network limiting user control. Finally, any benefit or information gained
from the middlebox is only effective while the device remains connected to that
network or to a particular location of a given network. Once the device leaves
and moves to a different network or location, it is difficult to re-connect the
traffic to the new location, especially in case of stateful services, where
state migration is required and particularly challenging in presence of large
number of highly mobile users.

\section{\m architecture}
\label{sec:arch}


In this section we present a brief description of \m design.
We start by discussing how we handle the traffic from apps on the device
before outlining how \m balances extensibility and security. 




\begin{figure}[t]
    \centering
    \hspace{20pt}
    \includegraphics[width=0.7\linewidth]{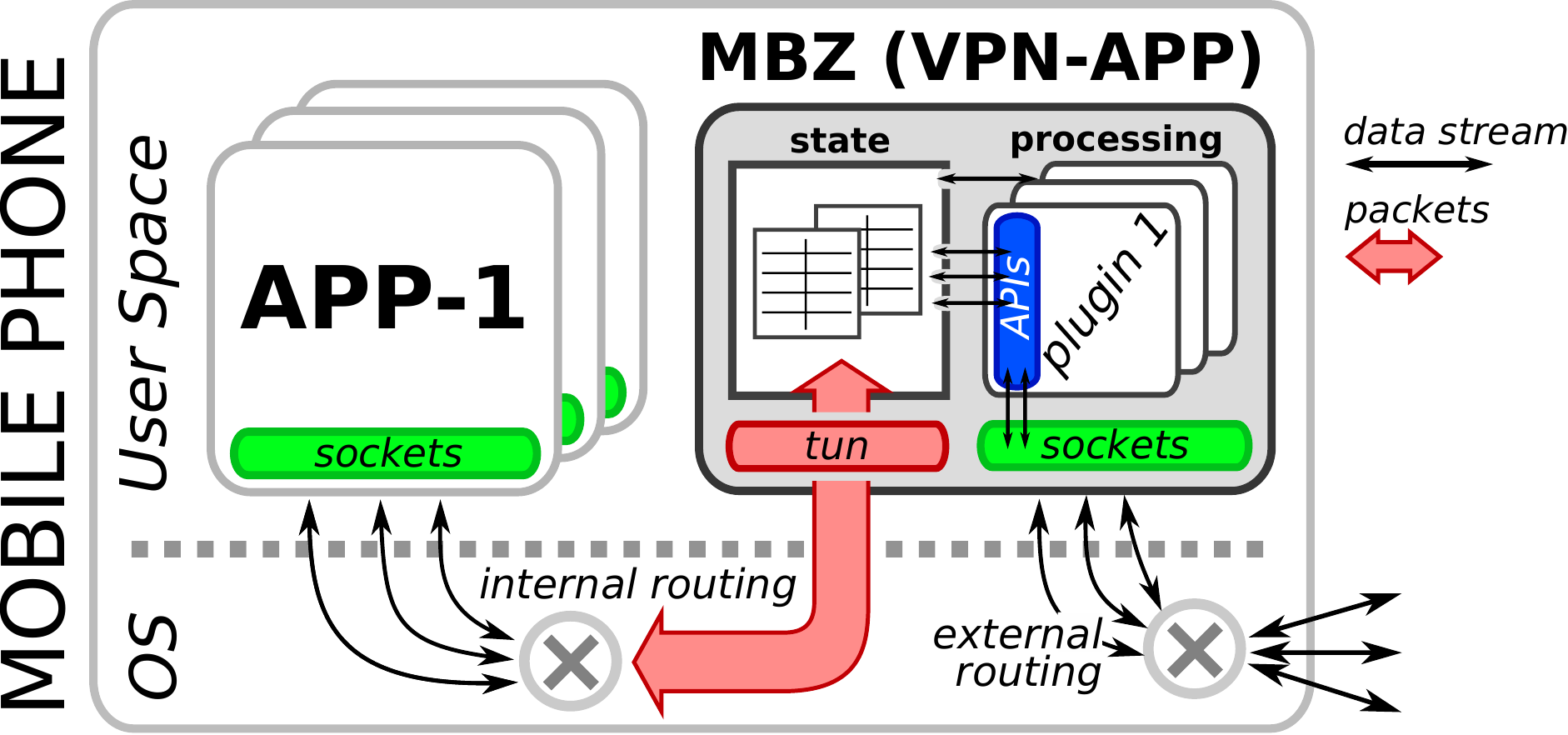}
    \caption{\footnotesize{Diagram of \m prototype.}} \label{fig:diagram}
    \vspace{-0.1in}\end{figure}

    
\subsection{Managing Traffic} \label{sec:monitor}

To operate as a middlebox, \m must be able to handle traffic before it leaves the 
device after being initiated by an application, and before it is passed to the application after being
received off the network. 

\m uses the VPN interface available as an API on today's
mobile phones. This approach has been used
before to detect privacy leaks~\cite{razaghpanah2015haystack} and monitor
network performance~\cite{wu2016mopeye,shuba2015antmonitor}. \m builds on it
with a much broader goal: to capture traffic and funnel it through
a series of user-specified extensions.\footnote{Thus, despite our discussion focused on mobile platforms, 
	any platform allowing the creation of virtual interface APIs could leverage an \m approach.}  

As depicted in Fig.~\ref{fig:diagram} the VPN API reveals a \tun interface that captures any 
traffic generated by applications on the device. While a typical VPN application
would forward the traffic to a server or proxy, \m handles that on the device. To 
accomplish this, the \m must extract flow state from the UDP/TCP and IP packet headers 
for packets arriving on the {\tt tun} interface and map the packet to a regular socket 
(creating sockets as needed). The \m must also maintain this state, both for UDP and TCP
flows, so that it can marshal data arriving from remote hosts on the
sockets back into packets for transmission to the app via the {\tt tun}
interface. 

In the case of TCP traffic, when the \m reads a SYN packet from the {\tt tun}
interface, it must create a new socket, connect to the endhost target and
instantiate state internally.  After the OS establishes the socket, the \m 
returns a SYN/ACK via the {\tt tun} interface to the originating app. 
Managing state  efficiently is critical, as a user-space \m is subject to the standard 
UNIX file descriptor limit and, so, must close sockets and flush state periodically.  
The explicit connection teardown of TCP connections provides a clear signal for
clean-up. However, for UDP's connectionless nature, we 
leverage inactivity timeouts as do NATs~\cite{richter2016imc}. One workaround 
would be to re-use sockets for UDP traffic, which can significantly reduce the 
number of sockets used for DNS resolutions, but does not solve the problem 
entirely.

\subsection{Plugin Management and Security}
\label{sec:plugins}

A key aspect of \m is its extensible nature. This is made possible by
a framework dedicated to installing and organizing plugins specified by
the user. We describe its preliminary design and workings in this section. 

We envision the plugins in \m to be similar to applications
installed on the device. There will be an ecosystem of plugins to be
installed and with the user having the ultimate say in which plugins are 
installed on in \m. In order for this to work efficiently and
safely, \m must ensure that the plugins cooperate to benefit the user
without interfering in or hindering the work of other plugins and the user
alike. 

There are two main factors that \m addresses when handling plugin
execution. First, plugins are granted a set of permissions to manipulate
user traffic similarly to browser extensions. \m ensures that every plugin
respects those permissions and no plugin abuses its power over the user's
traffic. There are several ways \m does this, first being that it limits
the actions that plugins are permitted to execute on packets. For
instance, unless specified by the user, \m prevents plugins from
collecting user data or routing traffic to a third-party server.

Second, \m guarantees resource isolation and management across plugins, similarly to 
existing NFV platforms.
Indeed, a processing/memory intensive plugin could consume all available resources 
hurting other applications or plugins performance and potentially impact user experience. 
Further, given the open nature of \m, it could be abused to initiate malicious activities. 
\m monitors plugins' resource usage and disables them if demands increase beyond a set of 
thresholds which could be set based on users' defined priorities and resources availability. 
Identifying and fine-tuning these thresholds remains future work. 
Further, we plan to make our resource management techniques context aware. For instance, 
it is critical to adapt resource allocation to battery levels, or adjust traffic generated 
by plugins to connectivity type (e.g., limit data exchange when a device is connected to cellular 
network to protect the user's data limits).




\section{Example Use Cases for \m}
\label{sec:apps}

The value of an on-board middlebox comes from combining the view of
the traffic that traditional middleboxes enjoy with the opportunity to
leverage the local---user-, device- and network-specific---context 
when processing traffic. 

We describe several use cases and applications for \m.
The list is clearly not exhaustive, but meant to illustrate 
some of the functionality that \m enables. As we discussed in \xref{sec:arch}, we
envision an architecture in which each use case operates as a separate
module or plugin to be installed by the user or by the app
developer. Some of these plugins could be bundled with \m as default
plugins, while others could be acquire from a third-party. 


\subsection{User-Defined Firewalls}
\label{sec:firewalls}

\begin{figure}[t]
    \centering
    \includegraphics[width=0.65\linewidth]{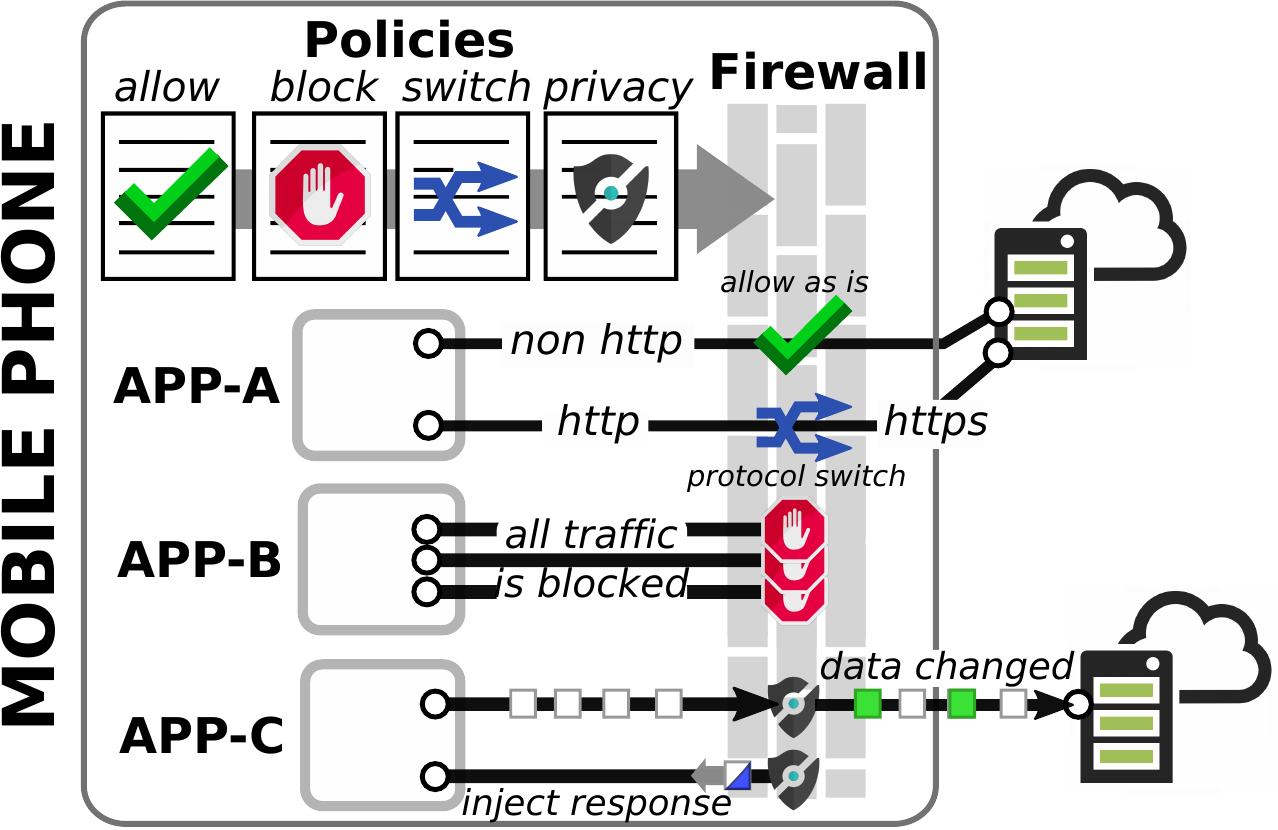}
    \caption{An \m extension could act as a firewall on the endhost, 
    	enacting user-set rules such as {\tt Switch}, or {\tt Deny}
        on the traffic.}
    \label{fig:firewall}
\end{figure}

Sending all traffic through \m enables much control over the
traffic, including the ability to monitor, block, redirect, shape
and change traffic in different ways.  While middleboxes within the
network have some of the same abilities, \m augments this ability with
device context which can enable more complex rules, especially because it
can identify which application is generating the traffic. For instance,
consider the following firewalling functions that \m can facilitate:

\parax{Blacklisting} The \m can implement blacklists (or
whitelists) that block (or enable) traffic to/from certain hosts or
domains, as in Fig.~\ref{fig:firewall}.  This can be done for all
traffic or just for certain apps.  For instance, one could
configure the \m to only allow the email application to interact with the
user's known IMAP and SMTP servers to thwart efforts to conduct email
tracking.

  
\parax{Protocol Usage} The \m can prohibit certain protocols in
specific instances or adjust protocol settings based on network conditions.  
For instance, the \m can ensure that a banking
app uses only encrypted connections.

\parax{Enhance Privacy} User tracking is omnipresent in the mobile
ecosystem \cite{ren2016recon,enck2010taintdroid,trackers:dat}.  The
\m is in a unique position to give users the ability to wrest back
control over their privacy by allowing them to either replace
private information inside network flows with random values before
they leave the device, or block such flows entirely.  These
controls can be precisely tuned using fine-grain details of the
communication, such as app, destination (\eg disallowing known
third-party trackers), protocol (\eg blocking unencrypted protocols), and
type of information being leaked.


The \m can use a variety of methods to enforce the smart
firewalling policies.  The exact method employed depends on the
specific use case and app pattern.  For instance, the \m could
prevent communication with a blacklisted host by issuing a TCP reset
packet to the app instead of instantiating a connection.
Alternatively, the \m could inject a response to a dubious request
letting the user know why their traffic was blocked.  This approach
could be taken a step further and the \m could alert users and
determine if the user would like to proceed anyway.  Another tool at
the \m's disposal is re-writing certain portions of the content to obscure
sensitive data (\eg re-writing the IMEI).  

\subsection{Fine-Grained Traffic Routing}

Given that the \m intercepts traffic before it leaves the device,
the system can re-direct the traffic based on users' preferences
rather than the current one-size-fits-all model of traffic
forwarding.

%

\parax{Smart Multihoming} Mobile phones often have multiple
connectivity options.  While traditionally phones only use
one network at a time, they are often in locales where multiple networks
are available (\eg cell and a WiFi). This opens
several opportunities.  For instance, users could identify their network
preference at some locations (such as WiFi at work), for all or specific apps
(WiFi at work, except for Facebook to skirt a policy block on the WiFi network).
Alternatively, the \m could monitor performance---see
\xref{sec:trouble}---and try to choose the ``best'' network at the given
time.

\parax{Smart Tunneling} Users often have the ability to connect to
myriad different private networks via tunnels.  For instance,
companies may force their employees to use VPNs to access content
hosted on their private network.  Or, privacy-aware individuals may
use commercial VPN services or anonymization networks such as Tor to
obfuscate the source of the traffic.  The \m can implement all these
policies.  Furthermore, \m can offer precise control such that a user's
work traffic is sent to their employer's VPN server while a particular
app's traffic may be directed to a VPN server hosted at user's home to
avoid geo-filters while traveling~\cite{choffnes2016case}.  This
capability of \m allows for unprecedented control instead of the current
one-size-fits-all approach.

\parax{Smart Multipathing} \m can also be used to enable
the use of multiple paths to aid performance.  This could be realized, for
instance, by turning normal TCP connections into a series of MP-TCP
subflows (which may or may not be divided across the device's network
interfaces). Alternatively, the \m could spread requests to multiple
instances of some replicated service (\eg a DNS resolver pool or among
edge servers in a CDN).
\vspace{-10pt}
\subsection{Network Troubleshooting}
\label{sec:trouble}

\m can observe the performance of all network traffic on the 
device, and thus can detect performance variations
and trends within a user's normal interactions, and contextualize 
those observations with link-level insights (\eg SNR). 

Allman and Paxson first described a reactive measurement approach
that advocates for considering measurement to be a process and not an
event \cite{allman2008reactive}.  While studies have used the
reactive measurement notion---\eg for broadband characterization
\cite{sanchez2013dasu}---\m provides a generic platform that enables
reactive measurements by, for instance, triggering active
measurements based on passive observations.

As an example, consider issues caused by the DNS resolution process.
A device could---likely in a sampled fashion---trigger alternate DNS
and resulting TCP transactions from passively observed traffic.  Now
the device can monitor multiple transactions that nominally do the
same task.  At this point we can engage in ``what if?'' analysis.
\Eg would performance be better or worse if hostnames were resolved
via a public DNS resolver (\eg OpenDNS or Google's Public DNS)?  \Eg
is the ISP manipulating the DNS responses in some fashion (to
monetize errors or for censorship)?  The answers to these questions
can then lead to configuration changes (\eg sending DNS queries
through a VPN to circumvent censorship).

\subsection{Improved Protocol Stacks}

Application developers can use a
variety of network- (\eg IPv4 or IPv6),
transport- (\eg UDP/QUIC or TCP) and application-layer (\eg HTTP(S)
or HTTP2) protocols, many interchangeable, to create mobile
applications. However, developers typically stick to a single
combination of application- and transport-level protocols for all
situations, regardless of network conditions. These limit the adaptability of applications in
today's evolving network conditions, particularly when a given
protocol or network element may outperform an equivalent one.


\m has the ability to experiment and examine how
different combinations of compatible
protocols perform over time for a given device and endhost 
machine, at a given time and network. 
This historical data can
be used by the \m to dynamically select a given application's protocol to
optimize application performance and ultimately the user experience.
Importantly, this is only possible because of \m's place in the network:
the end host. \m can implement changes and solutions that can help improve
the last mile of the network. Middleboxes within the network do not have
this ability.

One instance where a dynamic protocol stack could be beneficial to the
user is in scenarios with a high packet-loss and high latency in the last mile. 
The \m can identify when a network becomes lossy and adapt
the traffic to such adversary conditions. 

For example, \m can wrap underperforming TCP packets within a protocol
more tolerant to loss, such as QUIC~\cite{langley2017quic,rula2018mile},
either directly to the remote server or through a proxy. Likewise, the \m
can also prefetch DNS responses for the domains most relevant for the set
of applications running on the device, hence reducing the time required to
open a new connection by avoiding unnecessary DNS lookups. 

This last example also points to venues in which
\m could work in collaboration with middleboxes. This will allow as well  
\m to obtain information about the performance at the core of the network
(\eg congestion), data otherwise unavailable to endhosts.

\section{Information and Performance Validation}
\label{sec:validation}

In this section we present a proof of concept validation of \m.
First, we show \m can provide additional control and transparency to
the user with a sample {\it snitch} plugin. Second, we show the
limited overhead \m adds to the traffic. This is a simple demonstration of the potential
applications of \m and we leave a rigorous evaluation as future work.

\subsection{{\it What Third-parties are contacted by the website?}}
\label{sec:info}

\begin{figure}[t]
    \centering
    \includegraphics[width=0.60\linewidth]{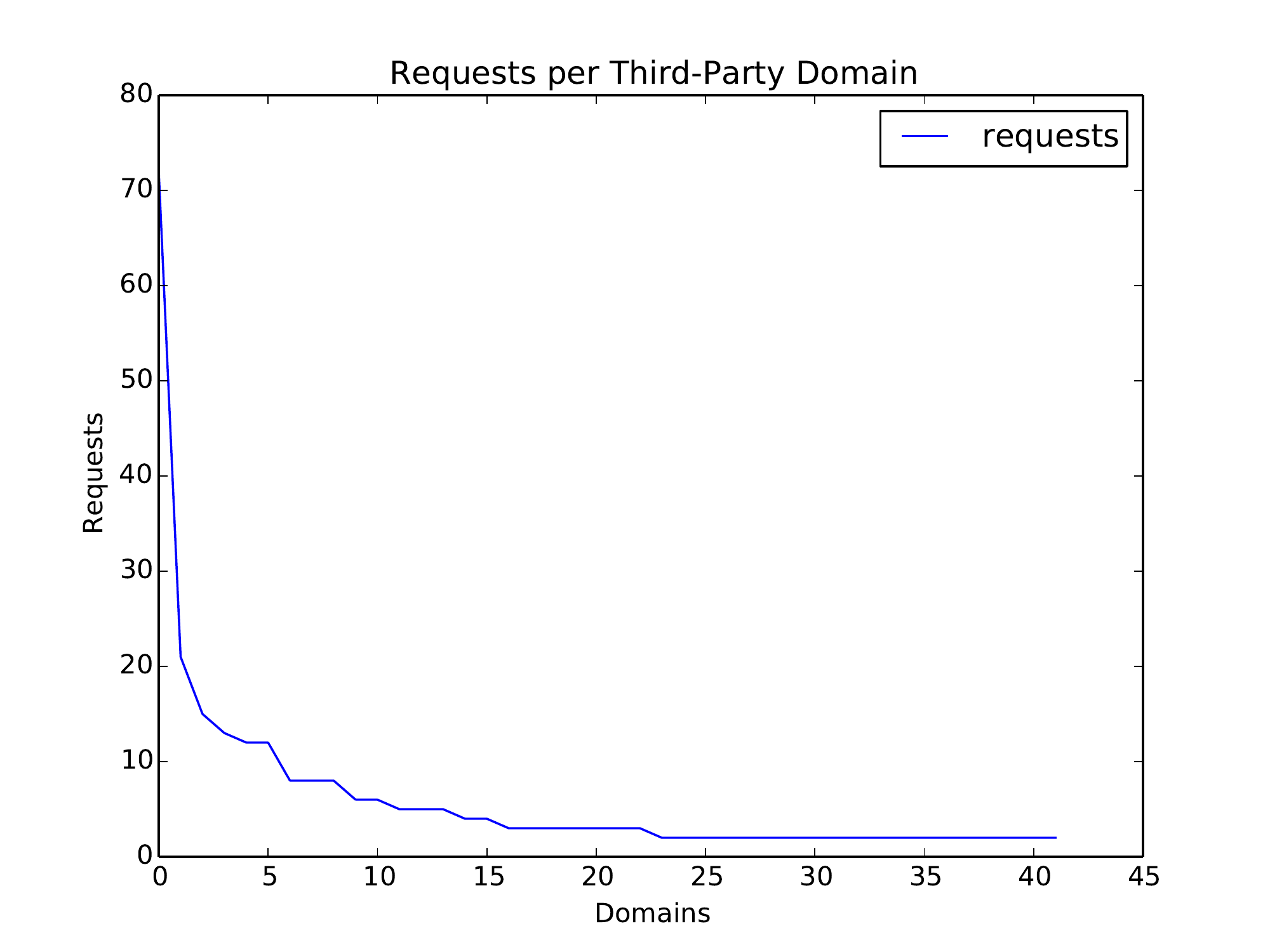}
    \caption{Number of requests  per third-party
    organization from real users running the Snapchat application. Data collected via \emph{snitch} MZB plugin in the wild.
    } \vspace{-.2in} \label{fig:reqs_per_dom} \end{figure}

We show how \m provides 
users access to information that would otherwise only be available by
rooting the phone or in a network middlebox. In this experiment, we set out to answer the question: {\it what
third-parties are contacted by the website?} We use an instantiation of \m
in the form of an Android application that has been deployed in the wild~\cite{razaghpanah2015haystack}.  
A basic {\it snitch} plugin is installed through \m to
investigate the traffic. The plugin passively monitors the connections
made by applications and tracks the destination IP address, port, and
protocol. This information is then displayed to the user in the \m
application. For lack of space, we report results for one popular application only: \emph{Snapchat}.


Figure~\ref{fig:reqs_per_dom} reports the number of requests generated towards different third-party domains. We observe that there are over 40
different organizations (i.e., third parties) contacted and, while most have only 1 request,
about 5 organizations account for more than 10 requests each. Additionally, \m can identify
the protocols used by different applications and share this information with the user. In the Snapchat case, we count a total of 372
third-party flows. The large majority of them are TCP flows (91.7\%)
while only 8.3\% are UDP. Interestingly, some of the UDP flows 
are leveraged to run the QUIC protocol, indicating that third-party organizations are
adopting Google's new protocol.

\subsection{Performance Evaluation}
\label{sec:measurement}

While it is possible to build \m using the Android VPN interface, it is
critical to ensure that it does not negatively affect the performance of
other applications. In this section we present a simple feasibility analysis of \m. 
To this purpose we use a rooted Android phone running {\tt tcpdump}~\footnote{\url{www.tcpdump.org}} which
would allow us to compare the end-to-end latency experienced by a sample
application with and without running our \m prototype. While working with
a non-optimized prototype, our analysis shows that there is minimal
overhead on the end-to-end latency of running applications. 

In our experiment, we use a test application that makes successive requests to each 
of the top 100 Alexa websites while measuring the connection time.
We compare the time from when the Java Socket API indicates a successful TCP connection 
with the TCP connection time on the network interface
according to {\tt tcpdump}, \ie the time between the first SYN and SYN/ACK. We first 
measure the TCP connection time without an \m running on the phone ({\em
no vpn}) as a baseline, so the extra latency is introduced by the  Java
Virtual Machine (JVM). This allows us to conclude that the JVM adds 77
microseconds (median) for every TCP connection. 

\begin{figure}[t]
    \centering
    \includegraphics[width=0.60\linewidth]{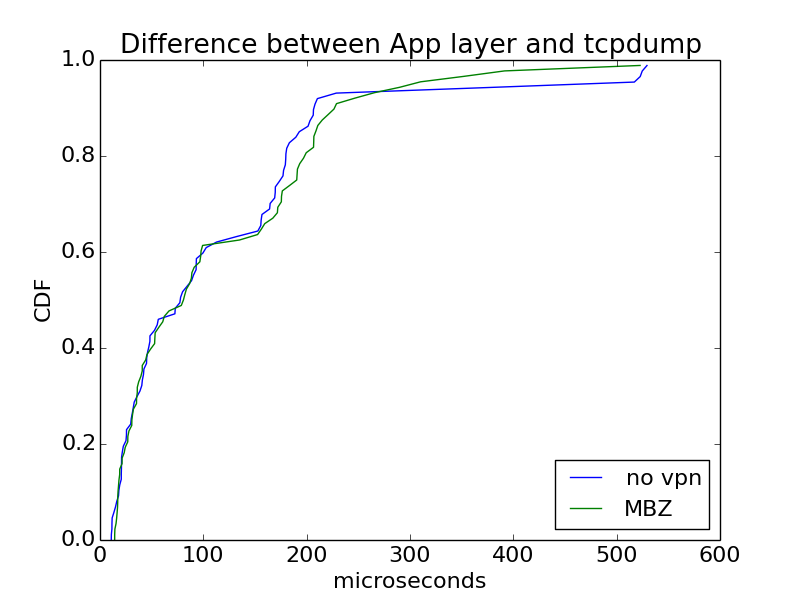}
	\caption{Difference in TCP connection time between tcpdump and 
        application level measurement.} \label{fig:app_level_cdf}
    \vspace{-0.1in} \end{figure}

	
To eliminate JVM's inherent latency, we implement a \m prototype in C++ ({\em \m}). 
Figure~\ref{fig:app_level_cdf} shows the difference between the time reported in 
{\tt tcpdump} and that on \m in microseconds. As the figure shows, the {\em \m}
curve closely overlaps that of the baseline ({\em no vpn}), mainly due to
C++'s ability to treat the {\tt tun} interface as a socket, making it
possible to poll the {\tt tun} interface along with the remote sockets,
which significantly reduces latency compared to alternating between
polling the {\tt tun} and the remote sockets.



\section{Conclusion}
\label{sec:conc}  

Users and researchers lack dynamic access and control over the
networking stack on modern end devices.  We advocate locating an extensible 
virtual middlebox on the end devices themselves. We call this approach 
``middlebox zero'' or \m. By being on-board, a \m also leverages local 
context as it processes the traffic and complement the network wide view 
of standard middleboxes. We discussed the challenges of the \m approach, 
sketch a working design, and illustrate its potential with some concrete 
examples that, as we argued, would be either cumbersome or not possible to 
realize using current solutions. We emphasize here that we are merely staking 
out a position in this paper. While we have developed a proof-of-concept 
implementation, many questions are left to resolve. Our goal with this paper is 
to share our ideas with the community in the hopes of gathering early feedback on
the \m approach.

\bibliographystyle{acm}
\bibliography{mbz}

\end{document}